%% file: sample-authordraft.tex
\documentclass[manuscript,revie=False]{acmart}
\AtBeginDocument{%
  \providecommand\BibTeX{{%
    \normalfont B\kern-0.5em{\scshape i\kern-0.25em b}\kern-0.8em\TeX}}}

\setcopyright{acmcopyright}
\copyrightyear{2018}
\acmYear{2018}
\acmDOI{XXXXXXX.XXXXXXX}

\acmConference[CHI EA ’24]{Make sure to enter the correct
  conference title from your rights confirmation emai}{May 11--16,
  2024}{ Hawai'i, USA}
\acmBooktitle{Extended Abstracts of the
2024 CHI Conference on Human Factors in Computing Systems (CHI EA ’24),
May 11–16, 2024, Hawai'i, USA} 
\acmPrice{15.00}
\acmISBN{978-1-4503-XXXX-X/18/06}

\usepackage{wasysym}

\begin{document}

\title{What is in Your App? Uncovering Privacy Risks of Female Health Applications}

\author{Muhammad Hassan}
\authornote{Both authors contributed equally to this research.}
\email{mhassa42@illinois.edu}
\orcid{0000-0002-5713-9658}
\author{Mahnoor Jameel}
\authornotemark[1]
\email{mjameel2@illinois.edu}
\affiliation{%
  \institution{University of Illinois Urbana-Champaign}
  \streetaddress{ 614 E. Daniel Street (the Hub)}
  \city{Champaign}
  \state{Illinois}
  \country{USA}
  \postcode{61820-6211}
}

\author{Tian Wang}
\affiliation{%
   \institution{University of Illinois Urbana-Champaign}
  \streetaddress{ 614 E. Daniel Street (the Hub)}
  \city{Champaign}
  \state{Illinois}
  \country{USA}
  \postcode{61820-6211}
}
\email{tianw7@illinois.edu}

\author{Masooda Bashir}
\affiliation{%
 \institution{University of Illinois Urbana-Champaign}
  \city{Champaign, IL}
  \country{USA}
}
\email{mnb@illinois.edu}

\renewcommand{\shortauthors}{Hassan and Jameel, et al.}

\begin{abstract}

    \emph{FemTech} or Female Technology, is an expanding field dedicated to providing affordable and accessible healthcare solutions for women, prominently through Female Health Applications that monitor health and reproductive data. With the leading app exceeding 1 billion downloads, these applications are gaining widespread popularity. However, amidst contemporary challenges to women's reproductive rights and privacy, there is a noticeable lack of comprehensive studies on the security and privacy aspects of these applications. This exploratory study delves into the privacy risks associated with seven popular applications. Our initial quantitative static analysis reveals varied and potentially risky permissions and numerous third-party trackers. Additionally, a preliminary examination of privacy policies indicates non-compliance with fundamental data privacy principles. These early findings highlight a critical gap in establishing robust privacy and security safeguards for FemTech apps, especially significant in a climate where women's reproductive rights face escalating threats.
    
\end{abstract}



\maketitle

\section{Introduction}
\label{sec:introduction}
\input{1_introduction}


\section{Background}
\label{sec:background}
\input{2_background}
\section{Methodology}
\label{sec:method}
\input{3_methodology}

\section{Results}
\label{sec:result}
\input{4_results}

\section{Conclusion}
\label{sec:conclusion}
\input{7_conclusion}

\bibliographystyle{ACM-Reference-Format}
\bibliography{sample-authordraft}

\appendix









\end{document}

%% file: 1_introduction.tex
In a rapidly developing digital age, technology constantly demonstrates itself to be an integral part of the lives of individuals. The efficiencies it provides allow for specific aspects of livelihood to be narrowed in on, targeted to fulfill specific needs, educate, and expand accessibility. This rings true especially in the case of female. From well-known tools such as Flo Health, Maven Clinic, and Progyny where specific emphasis is placed on reproductive health, technology has placed itself in a role that promotes the empowerment and agency of female health\cite{Narwani_2023, Bose_2023}. 

Tools such as Flo Health and Progyny are in the technological category of FemTech, a term that refers to the use of technology to address the needs and challenges of women’s health and wellness \cite{yasharfuture, Bose_2023}. FemTech mobile apps are one of the most popular and accessible forms of this technology, offering a variety of features and services for women, such as period tracking, fertility and contraception support, menopause management, hormonal disorder and chronic condition education, and sexual wellness guidance. Provided technology’s strong, deep presence and design for education, personal monitoring, and solution-based recommendations, one must consider the impacts and any related considerations around the vulnerabilities that exist and the biggest ones being privacy and policy.

The recent reversal of Roe v. Wade by the US Supreme Court, which recognized a woman’s constitutional right to choose whether to have an abortion, has raised serious concerns about the reproductive rights and privacy of women in the United States\cite{Totenberg_McCammon_2022}. The decision has not only banned or restricted access to safe and legal abortions, but also undermined the right to privacy that was the basis of Roe v. Wade, which could have implications for other aspects of women’s health care such as fertility treatment, contraception, and cancer care \cite{Scott_2022, suran2022treating}. Moreover, there have been cases where women have been prosecuted or persecuted based on their online purchase of abortifacients, text messages, or search history related to abortion or miscarriage\cite{Valenti_2015, Elliott_2022, antyour}. Therefore, some experts have advised women to delete their FemTech apps or avoid using them to track their periods and fertility cycles, as they fear that their personal data could be used against them by anti-abortion laws or authorities\cite{antyour}.

Female health applications (FHA) are a form of FemTech that provide various services and products to women, such as period tracking, fertility support, pregnancy monitoring, menopause management and general health care. However, these applications also collect sensitive health data from their users, which could expose them to privacy and security risks. Previous studies have shown that some FHA have poor data security and privacy practices, such as tracking and profiling users, sharing data with third parties, or not disclosing data policies to users \cite{wiredWIREDGuide, shipp2020private, alfawzan2022privacy, mehrnezhad2021caring}. Moreover, in some contexts, women’s bodies are under political surveillance, and their data could be used against them by authorities or adversaries. For example, in Missouri, the state health department tracked women’s period information to determine if an abortion took place \cite{theguardianGovernmentTracks}. Therefore, it is important to understand the security, privacy and data practices of FHA and how they affect the users. In this exploratory study, we aim to provide a comprehensive overview of the current state of FemTech apps and examine their potential benefits and challenges for women’s health. This emphasis on data privacy informs underlying impacts on threats to female health posed by data vulnerabilities. We conduct a preliminary analysis of what and how information is collected by FHA, and what information is disclosed to the users about the data practices of FHA by their privacy policies.

We present an initial exploration of the FHA space that aims to answer the following research questions.

\begin{itemize}

    \item \textbf{\emph{RQ1} Examining the Data collection and data practices of popular FHA?} \\ We analyzed the permissions requested by the applications to determine their level of privilege and access to sensitive data. We also examined the presence of third-party trackers in the applications that may collect and share user data.

    \item \textbf{\emph{RQ2} Evaluating privacy policies of selected FHA apps for data privacy principles} \\  We assessed the privacy policies of 7 popular FHA apps by using the Fair Information Practice Principles FIPPs framework to evaluate their data practices focusing on privacy and security measures.
    
\end{itemize}

%% file: 2_background.tex
The academic literature on the use of technology for female health \emph{(FemTech)} reveals the importance and the risks of collecting and processing female healthcare data. In this section, we review the relevant literature to understand the current state of FemTech and the challenges it faces in terms of data privacy and security. We also examine the laws and policies that regulate healthcare data in general and female healthcare data in particular, focusing on prominent frameworks in this domain. We compare and contrast these regulations to investigate how they address the specific needs and concerns of female healthcare data privacy. Our goal is to identify the gaps and opportunities for improving FemTech data privacy and security in light of the existing literature and regulations.

\subsection{Female Health Data Risks}

Female Health Apps (FHA) collect sensitive user data, often revealing intimate aspects of women’s lives. Trust and data safety are paramount, especially for FHA users, as they entrust these applications with their sexual and reproductive health (SRH) information\cite{aimeur2016changing, kesan2015comprehensive, muller2023you}. However, uninformed users may share data without understanding the implications\cite{uuInformedConsent,vallor2018introduction, umic_informedconsent}. Previous studies highlight the privacy risks in FemTech and FHA due to inadequate consent or protection\cite{almeida2022bodies}, and the misuse of this data could lead to discrimination, violence, and legal repercussions\cite{vanderbiltDataPrivacy, iappPrivacyDigital, almeida2022bodies, theguardianGovernmentTracks, justiaDobbsJackson, constitutioncenterDobbsJackson, reproductiverightsCaseDepth, morganlewisEvolvingLaws}.

FHA’s collection of personal health data poses privacy risks. Sharing this data with third parties could lead to unwanted targeting or exploitation. Despite privacy promises, some FHA have been found to share user data with third parties\cite{consumerreportsWhatYour, consumerreportsReproductiveHealth}, and some even shared health information without user consent\cite{ftcLocationHealth}, breaching trust and privacy rights.

%% file: 3_methodology.tex
To begin, we select seven women's health applications and conduct a static analysis of the applications' code to reveal their functionalities. Next, we analyze the privacy policies of these applications to determine their scope, accessibility, and ease of understanding.

Our research employs a mixed-method approach, combining both qualitative and quantitative analyses, to address a range of issues related to women's health applications. This approach is essential as it allows us to investigate both the linguistic analysis of privacy policies and the technical behavior of the applications. This section outlines the various steps we undertook in our research methodology.

\subsection{App Selection}

\input{tables/apps}

Our study focuses on analyzing Android Google Playstore applications, which are widely available and offer many resources for analysis. To identify relevant applications, we use search terms such as \emph{'woman', 'female', 'feminine'} along with combination of \emph{'health', 'wellness', and 'well-being'} on the Google Playstore. We rank the search results based on the number of downloads and ratings and downloaded the applications using a new Google Account in May 2023. Our selection of applications follows the research methodology of previous studies such as Adhikari et al. \cite{adhikari2014security} and Shipp et al. \cite{shipp2020private}. During our analysis, we exclude applications that were unrelated to our research objective, such as general awareness applications that appeared in the search results due to associated tags or descriptions. Additionally, we excluded \emph{'erotica'} applications as they were irrelevant to the functionality we aimed to investigate.

\subsection{Static Analysis}
In this step of our analysis, we employed a rigorous approach to examine the applications by installing them on an Android device and using Android Debugging Bridge (adb) to extract the apk file onto a workstation. To conduct the static analysis of the applications, we utilized Mobile Security Framework \cite{abraham2016mobile}, which is a widely adopted tool for this purpose, and in accordance with the methodology of \cite{owens2022electronic}. The application's source code contains vital information about its data practices, including the permissions it requests, which act as data sources and may lead to privilege escalation \cite{felt2011android, li2021android}. 
Google groups permissions into 4 categories based on the risk associated with the permission and the use\cite{androidpermissionAndroid}. These permission categories include 
\begin{itemize}

    \item \textbf{Normal Permissions} are considered low-risk permissions for system and other applications.
    \item  \textbf{Dangerous Permissions} are higher-risk permissions granting an application access to private user data or control over the device.
    \item \textbf{Signature Permissions} are granted automatically to an application if it is signed with the same certificate as the application that declared the permission, without informing or approval of the user.
    \item \textbf{signatureOrSystem} are used for sharing specific features between multiple vendors’ applications built into a system image.

\end{itemize}

To gain deeper insights into the applications' data practices, we also examined the presence of third-party libraries and trackers within the application code. Developers often integrate these third-party services to monetize, analyze, or add new functionalities to their applications. However, these libraries or third parties may pose a threat to user privacy by leaking users' information \cite{balebako2014privacy}. Therefore, the presence of a higher number of permissions declared by an application may potentially escalate its privilege, and the presence of a higher number of trackers may indicate a privacy vulnerability.

\subsection{Privacy Policy Analysis}
In this phase of our analysis, we collected the privacy policies of the seven selected applications. It is ideal for the privacy policy to be readily available on the app page in Google Playstore; however, if not, we ensured that it was available after downloading the application. We used the Fair Information Practice Principles (FIPPs) as a standard to evaluate the privacy policies of the selected applications \cite{fpcFPCgov}. FIPPs are a set of established principles that form the foundation for both GDPR and HIPAA. In addition to the evaluation based on FIPPs, we also analyzed the privacy policies to identify the user ('data subject') rights listed in them. This analysis provides insight into the transparency and accountability of the application developers regarding the handling of user data.

%% file: tables/apps.tex
\begin{table}
  
    \caption{List of the Applications Analyzed}
    \label{tab:app_selection}
    \begin{tabular}{ccl}
    \toprule
    \hline 
    \textbf{App \#}&\textbf{Installs} &\textbf{Category}  \\
    \midrule
    \hline 

1   & 1,000,000,000+    & Health and Fitness    \\ \hline
2   & 100,000,000+      & Health and Fitness    \\ \hline
3   & 100,000,000+      & Health and Fitness    \\ \hline
4   & 10,000,000+       & Health and Fitness    \\ \hline
5   & 500,000+          & Medical                \\ \hline 
6   & 100,000+          & Health and Fitness    \\ \hline 
7   & 100,000+          & Health and Fitness    \\ \hline 
  \bottomrule
\end{tabular}
\end{table}

%% file: 4_results.tex
In order to comprehensively analyze the selected women's health applications, we conducted a thorough examination of their permissions, privacy policies, and third-party library usage. In this sections, we will present our findings for all the applications included in our study.

\subsection{Permission: Information Source}

\begin{figure}[t]
\includegraphics[width=14cm]{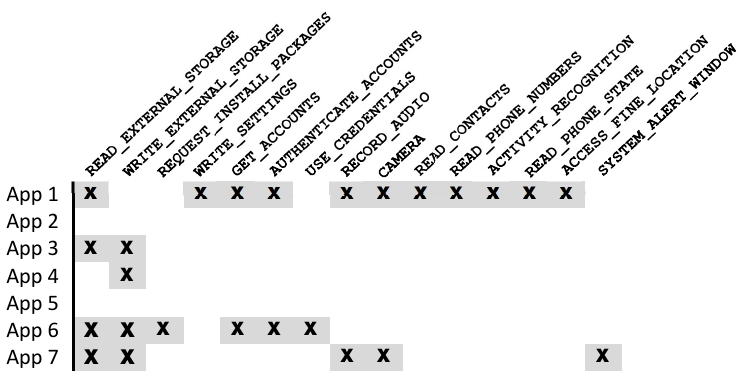}
\caption{Dangerous Permissions in Applications}
\end{figure}

The scope of data that can be collected by apps is largely determined by the permissions they request. To thoroughly evaluate the privacy risks to individuals using these apps, it is essential to understand the types of data that can be potentially collected. The permissions on smartphone operating systems are in place to safeguard restricted data and actions. Consequently, apps that request more permissions have the more privilege to transmit a larger amount of data to backend and third-party entities. Analyzing the distribution of permissions requested by these female health apps allows for comparison to the app with the least amount of privileges among them. If the app with the least permissions shares similar goals with the others, it may function as a standard for the "minimum number of permissions" required for these apps to operate optimally. This method is in accordance with the Owens. et. al. approach for permission analysis. \cite{owens2022electronic}

\textbf{The most prevalent dangerous permissions gave applications access to external storage.} \verb|READ_EXTERNAL_STORAGE| and \verb|WRITE_EXTERNAL_STORAGE| were most common permission found in the study. We found four (4/7) application to use these permissions. These permissions allow applications to read and write data to external storage devices, potentially exposing sensitive user data to unauthorized access.


\textbf{We observed several requested permissions that were not commonly used.} One such permission is \verb|REQUEST_INSTALL_PACKAGES|, which, if granted, can be used to install any app package on the user’s device \cite{googleREQUESTINSTALLPACKAGESPermission}. Another permission, \verb|WRITE_SETTINGS|, was requested by an application and can be used to change a system setting of the device. The permissions \verb|READ_CONTACTS| and \verb|READ_PHONE_NUMBERS| were also requested once and grant apps access to the user’s contact information and device phone number(s), respectively. The \verb|READ_PHONE_STATE| permission was requested by an application and grants access to the phone state, including the current cellular network information, the status of any ongoing calls, and a list of any Phone Accounts registered on the device. This is a highly privileged permission that includes the \verb|READ_PHONE_NUMBERS| permission and the \verb|PhoneAccount| class as a subset \cite{androidManifestpermissionAndroidREAD_PHONE_STATE}.

The permissions \verb|ACCESS_FINE_LOCATION| and \verb|ACTIVITY_RECOGNITION| were requested once and grant app access to the user’s fine location and physical activity, such as running, walking, biking etc., respectively \cite{androidManifestpermissionAndroidACCESS_FINE_LOCATION, androidManifestpermissionAndroidACTIVITY_RECOGNITION}. Lastly, the \verb|SYSTEM_ALERT_WINDOW| permission was also only requested once and allows an app to create windows that are displayed over all other apps, such as pop-up notifications or floating widgets etc. Google recommends that this permission be used by very few applications as the  \emph{windows} created are for system-level interactions \cite{androidManifestpermissionAndroidSYSTEM_ALERT_WINDOW}.

\subsubsection{Most Privileged App}
Our analysis of mobile applications and their data practices revealed that \emph{App 1} has the highest level of privilege among the applications in our study. This is because it requests a large number of dangerous permissions, which grant access to sensitive data and resources on the device. Moreover, we found that App 1 requests most of the least widely used dangerous permissions, which are uncommon among other applications. Considering that App 1 has over 1 billion downloads on the Google Play Store, its data practices have a significant impact on a large number of Android users. We are currently conducting a further study to examine the functionality and features of App 1 and to evaluate the necessity and justification of its permission requests.

\subsubsection{Least Privileged App(s)}

In contrast to App 1, our analysis revealed that two other health applications, \emph{App 2} and \emph{App 5}, did not request any dangerous permissions. Despite this difference in data practices, both applications have a significant user base. App 2 has over 100 million downloads, while App 5 has over 500K downloads.

\subsection{Trackers: Information Sink}

As shown in Table \ref{tab:trackers}, our analysis detected the presence of various trackers within the applications studied. The majority of these trackers were utilized for analytic and advertisement purposes. Notably, Facebook had the highest number of unique trackers present (5), which served a range of functions including analytics (based on user interaction data within the app), user identification (for login and sign-up purposes), and content sharing. Google also had a significant presence with 3 trackers, primarily focused on analytics and advertisement.

\textbf{Our analysis revealed that the highest number of trackers present in a single app was 8.} Specifically, the app \emph{App 4} contained 8 unique third-party trackers, including those from both Facebook and Google. The primary functions of these trackers were for advertisement and marketing purposes, as well as for analytic based on user interaction with the application. The app with the second-highest number of trackers was  \emph{App 3}, with 7 trackers present. These trackers served similar functions to those found in App 4.

\input{tables/trackers}

\subsection{Privacy Policy: Information Notice}

As part of our analysis of privacy policies, we evaluated the availability of privacy policies on the Google Play Store page for each application. Ideally, an app should provide its privacy policy on its Play Store page for easy accessibility. Our findings revealed that approximately 40\% applications in our study did not provide a privacy policy on their Play Store page. These application account for more than 20 Million download, hence it takes privacy accessibility away from a large number of users.

\textbf{Our analysis of the privacy policies of the selected applications revealed that none of them adhered to the principles of \emph{Data Minimization} and \emph{Quality and Integrity}.} We found that these apps’ privacy policies claimed to collect more personally identifiable information (PII), such as user information and device identifiers, than was necessary for their core functionality. Furthermore, these apps did not limit their retention of PII and user data. Instead, they would either claim to retain it for business purposes even after a user deleted their account, keep it for an extended period of time, or use unclear and vague language regarding data retention. Some applications did not provide any notice on how long data would be kept. Additionally, none of the applications provided any notice on how they would ensure the accuracy and relevance of the data they collected to ensure fairness in their services.

\textbf{The analysis further reveals that the least followed principles by the apps’ privacy policies were \emph{Accountability}, \emph{Authority}, and \emph{Purpose Specification and Use Limitation}}. We found that these apps’ privacy policies did not provide clear information on who is responsible for ensuring compliance with privacy regulations, who has the authority to access and use user data, and whether any training is provided to those who access users’ data and PII. Furthermore, there was a general lack of detail regarding the specific purposes for which user data is collected and used, which also correlates with the lack of adherence to the principle of \emph{Data Minimization}. As a result, users may not have a clear understanding of how their data is being collected, used, and protected by these apps. 

\textbf{A majority of applications (70\%) were observed following principle of \emph{Individual Participation}.} Our analysis of the privacy policies of the applications under study revealed that they purport to involve user consent in their data practices. Specifically, these policies provide mechanisms for users to submit complaints or share concerns and queries regarding privacy and data processing. This indicates that the privacy policies claim to take measures to ensure that users have a degree of agency in providing input or feedback on personal data processing and can participate in decisions regarding its collection and use.

\input{tables/policies}

%% file: tables/trackers.tex
\begin{table}
  
    \caption{Third party trackers detected in the applications}
    \label{tab:trackers}
    \begin{tabular}{lcl}
    \toprule
    \hline 
    \textbf{Trackers}&\textbf{\# of Apps} &\textbf{Category}  \\
    \midrule
    \hline 
    AppsFlyer                 & 2                                                                & Analytics                         \\ \hline
AutoNavi / Amap           & 1                                                                & Location                          \\ \hline
Facebook Ads              & 3                                                                & Advertisement                     \\ \hline
Facebook Analytics        & 2                                                                & Analytics                         \\ \hline
Facebook Login            & 3                                                                & Identification                    \\ \hline
Facebook Places           & 2                                                                &                                   \\ \hline
Facebook Share            & 3                                                                & Content Sharing                   \\ \hline
Google AdMob              & 3                                                                & Advertisement                     \\ \hline
Google CrashLytics        & 3                                                                & Crash Reporting                   \\ \hline
Google Firebase Analytics & 5                                                                & Analytics \\ \hline
IAB Open Measurement      & 1                                                                & \begin{tabular}[c]{@{}l@{}} Identification \\ \& Advertisement \end{tabular}    \\ \hline
myTarget                  & 1                                                                & Advertisement                     \\ \hline
myTracker                 & 1                                                                & \begin{tabular}[c]{@{}l@{}} Analytics \\ \& Marketing    \end{tabular}    \\ \hline
OneSignal                 & 1                                                                & \begin{tabular}[c]{@{}l@{}} Push Notification \\ \& Messaging \end{tabular}  \\ \hline
  \bottomrule
\end{tabular}
\end{table}

%% file: tables/policies.tex
\begin{table*}[]
  
    \label{tab:policies}
    \centering
    \caption{Policies Analysis of Female Health Application (LEGEND: \CIRCLE = Followed, \RIGHTcircle = Partial, \Circle = Not followed)}
    \begin{tabular}{|l|c|c|c|c|c|c|c|}
    \toprule
    \textbf{Privacy principles} & \textbf{{App 1}} & \textbf{{App 2}} & \textbf{{App 3}} & \textbf{{App 4}} & \textbf{{App 5}} & \textbf{{App 6}} & \textbf{{App 7}} \\ 
    \midrule
    \hline 

\begin{tabular}[c]{@{}l@{}}\textbf {1. Access and Rectification}\\ \end{tabular} & \Circle &  \CIRCLE & \Circle & \Circle & \Circle &  \CIRCLE & \Circle \\ \hline
\begin{tabular}[c]{@{}l@{}}\textbf{2. Accountability}\\ \end{tabular} & \Circle & \RIGHTcircle & \Circle & \Circle & \Circle & \Circle & \Circle \\ \hline
\begin{tabular}[c]{@{}l@{}}\textbf{3. Authority}\\ \end{tabular} & \Circle &  \CIRCLE & \Circle & \Circle & \Circle & \Circle & \Circle \\ \hline
\begin{tabular}[c]{@{}l@{}}\textbf{4. Minimization}\\\end{tabular} &\Circle  &\Circle  & \Circle & \Circle & \Circle &\Circle  &  \Circle\\ \hline
\begin{tabular}[c]{@{}l@{}}\textbf{5. Quality and Integrity}\\ \end{tabular} & \Circle & \Circle &  \Circle & \Circle & \Circle & \Circle & \Circle \\ \hline
\begin{tabular}[c]{@{}l@{}}\textbf{6. Individual Participation}\\\end{tabular} & \RIGHTcircle &  \CIRCLE & \RIGHTcircle & \Circle & \RIGHTcircle & \RIGHTcircle & \Circle \\ \hline
\begin{tabular}[c]{@{}l@{}}\textbf{7. Purpose Specification and Use Limitation}\\ \end{tabular} & \Circle &  \CIRCLE & \Circle & \Circle & \Circle & \Circle  &  \Circle \\ \hline
\begin{tabular}[c]{@{}l@{}}\textbf{8. Security}\\ \end{tabular} & \Circle &  \CIRCLE & \RIGHTcircle & \Circle & \Circle & \Circle & \Circle \\ \hline
\begin{tabular}[c]{@{}l@{}}\textbf{9. Transparency}\\ \end{tabular} & \Circle & \RIGHTcircle & \Circle & \Circle & \RIGHTcircle & \Circle & \Circle \\ \hline

\end{tabular}

\centering
\end{table*}

%% file: 7_conclusion.tex
In this paper, we conducted an exploratory study on the privacy and security of 7 popular Female Health Applications. We found that these applications requested a varied number of dangerous permissions, which gave them access to sensitive data and resources on the device. We also detected numerous third-party trackers in these applications, which could collect and share user data with external parties, such as advertisers, analytics providers, or social media platforms. Furthermore, we analyzed the privacy policies of these applications using the FIPPs framework and found a general lack of adherence to various principles, such as notice, choice, access, security, and accountability. These findings raise concerns about the privacy and security of user information, especially in the context of the current political and capital surveillance of female health data in the post Roe v. Wade era. To the best of our knowledge, this is the first comprehensive study on the overall female health applications, extending the limited prior work on sub-categories. We are currently working on an extended set of applications with an enhanced analysis pipeline for a further study on this topic. We hope that our current and future results will help all stakeholders improve privacy design by ensuring more informed user policies and data privacy practices.
